\documentclass[aps,prl,preprint,groupedaddress]{revtex4}

\usepackage{graphicx,epsfig,amsmath,bbold}

\newcommand{\myvec}[1]{\boldsymbol{#1}}
\newcommand{\myten}[1]{\mathsf{#1}}

\newcommand{\mscl}[0]{\textsc{m}sc\textsc{l}}
\newcommand{\mscs}[0]{\textsc{m}sc\textsc{s}}
\newcommand{\kT}[0]{k_{\textrm{B}}T}
\newcommand{\pmf}[0]{\textsc{pmf}}

\begin{document}

\title{Mechanosensitive Channel Activation by Diffusio-Osmotic Force}

\author{Douwe Jan Bonthuis}
\email[]{douwe.bonthuis@physics.ox.ac.uk}
\thanks{corresponding author}
\author{Ramin Golestanian}
\email[]{ramin.golestanian@physics.ox.ac.uk}
\thanks{corresponding author}
\affiliation{Rudolf Peierls Centre for Theoretical Physics, University of Oxford, Oxford OX1 3NP, United Kingdom}

\date{\today}

\begin{abstract}
For ion channel gating, the appearance of two distinct conformational states and the discrete transitions between them is essential, and therefore of crucial importance to all living organisms.
We show that the physical interplay between two structural elements that are commonly present in bacterial mechanosensitive channels, namely a charged vestibule and a hydrophobic constriction, creates two distinct conformational states, open and closed, as well as the gating between them.
We solve the nonequilibrium Stokes-Poisson-Nernst-Planck equations, extended to include a molecular potential of mean force, and show that a first order transition between the closed and open states arises naturally from the diffusio-osmotic stress caused by the ions and water inside the channel and the elastic restoring force from the membrane.
Our proposed gating mechanism is likely to be important for a broad range of ion channels,
as well as for biomimetic channels and ion channel-targeting therapeutics.
\end{abstract}

\maketitle

Osmotic shock presents a fatal risk to unicellular organisms.
A sudden increase of the environmental solute concentration, known as hypertonic shock, leads to water loss and cell volume decline, whereas a sudden decrease, referred to as hypotonic shock, causes water to enter the cell rapidly, inducing cytolysis.
As a final resort in case of severe hypotonic shock, many bacteria, archaea and fungi avert cell lysis by activating non-selective membrane channels to release solutes from the cytoplasm \cite{2010_Kung}.
In \textit{E. coli} bacteria, two well-studied membrane protein channels are responsible for the release of solutes: the mechanosensitive channel of large conductance (\mscl{}) \cite{2002_Perozo_Nature} and the mechanosensitive channel of small conductance (\mscs{}) \cite{2008_Vasquez}.
Based on the observation that mechanosensitive channels are activated \textit{in vitro} by an applied hydrostatic pressure, the prevalent model for the gating mechanism invokes a conformational change in the protein triggered by tension applied to the cell membrane \cite{1994_Sukharev_Nature, 2002_Perozo_Nature, 2008_Vasquez}.
A free energy landscape for channel activation can be constructed by considering an elastic force proportional to the applied pressure \cite{2004_Wiggins-Phillips}.
However, no proposed gating hypothesis has been able to explain the appearance of two distinct conformational states and the discrete transitions between them.
In \textit{E. coli} \mscl{} mutants, added charge in the pore region activates the channels also in the absence of a hydrostatic pressure difference \cite{2001_Yoshimura,2004_Bartlett, 2006_Bartlett, 2002_Batiza}, highlighting the importance of electrostatic interactions in the activation process.
Indeed, the transmembrane domains of both \mscl{} and \mscs{} carry a substantial net charge: Each of the ten transmembrane helices of the pentameric \mscl{} protein carries a net charge of $-1e$ \cite{1998_Gu-Liu-Martinac}, and the heptameric \mscs{} protein carries an arginine residue with a charge of $+1e$ on each of its monomers \cite{2006_Sotomayor}.
Charge-induced activation is a robust feature of \mscl{} channels and has been used for drug delivery into mammalian cells \cite{2012_Doerner}.
Despite its significance, however, the electrostatic contribution to the activation energy, and in particular the diffusio-osmotic force originating in the dynamic overlapping double layer at the channel's charged surface, has not been considered up to now.

\begin{figure}
\includegraphics[width=8cm]{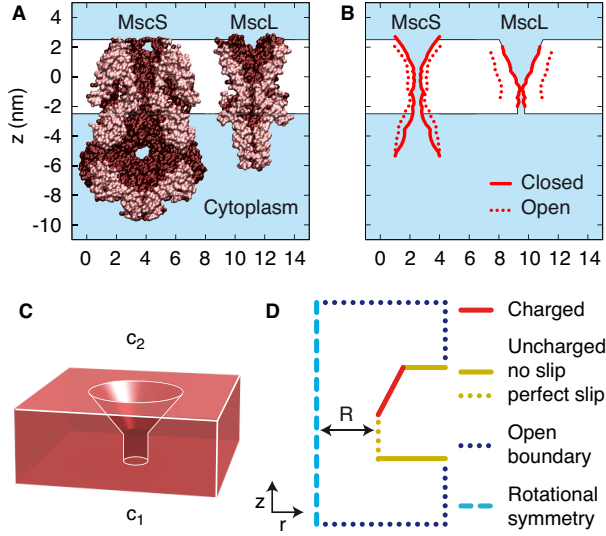}
\caption{
(A) Cross sections of the crystal structures of the mechanosensitive channels of large conductance, \mscl{} (2OAR), and small conductance, \mscs{} (2OAU).
(B) Pore outlines of the closed and open configurations of \mscl{} \cite{2002_Perozo_Nature} and \mscs{} \cite{2008_Vasquez}.
(C) Sketch of the channel embedded in a section of the membrane.
(D) Computational domain in cylindrical coordinates with the boundary conditions used \cite{supplement}.
}
\label{fig:boundary-conditions}
\end{figure}

The permeation pathways of both \mscl{} and \mscs{} are funnel-shaped, with the conical vestibule opening to the periplasmic side \cite{2002_Perozo_Nature, 2008_Vasquez}, and the stem of the funnel lined with uncharged hydrophobic residues (Fig. \ref{fig:boundary-conditions}A).
Upon activation the pore walls move radially outward (Fig. \ref{fig:boundary-conditions}B).
In weakly polar channels, water can fill constrictions down to the size of a single water molecule \cite{2004_Beckstein},
but even strongly hydrophobic channels are intermittently filled with water \cite{2001_Hummer_Nature, 2004_Beckstein, 2002_Allen_PRL}.
Ions, on the other hand, are subject to a strongly repulsive potential of mean force (\pmf{}) up to channel radii much larger than the ionic radius, caused by their hydration shells \cite{2012_Hummer, 2012_Richards, 2004_Allen_PNAS}, steric and van der Waals interactions and self energy \cite{1969_Parsegian}.
Using molecular dynamics simulations, the energy barrier for ion permeation through \mscs{} has been estimated to be $17$--$34$~$\kT{}$ \cite{2004_Anishkin}, explaining the lack of electric conductivity of \mscs{} in the closed state despite its relatively wide permeation pathway.
A similar hydrophobic lock mechanism has been found in \mscl{} \cite{2002_Perozo_Nature} and many different membrane channels \cite{2003_Kuo_Science, 2003_Miyazawa, 2011_Payandeh_Nature}.
Using mutational analysis, it has been established that the hydrophobic constriction in \mscl{} provides a threshold for channel activation \cite{1999_Yoshimura}.

The use of continuum hydrodynamics in nanometer-sized tubes has been shown to be justified for radii in the nanometer range \cite{2009_Thomas-McGaughey};
a noteworthy result, which can be rationalized by analytic arguments \cite{2010_Bocquet_ChemSocRev} and has been used recently to calculate the hydrodynamic resistance of aquaporin channels \cite{2013_Gravelle_PNAS}.
Similarly, the Nernst-Planck equation for ion transport has been found to be applicable down to a radius of $R = 0.3$~nm, provided that the ion concentrations are estimated accurately \cite{2011_Song-Corry}.
Ion concentrations at solid surfaces and lipid bilayers can be accurately calculated from mean-field theory when the ionic \pmf{}, estimated using molecular dynamics simulations, is included as a non-electrostatic contribution to the potential \cite{2008_Horinek, 2013_Bonthuis_JPCB}.
Combined, extended mean-field theory and continuum hydrodynamics reproduce the electrokinetic properties found in experiments and atomistic simulations of both hydrophilic and hydrophobic surfaces \cite{2012_Bonthuis_Langmuir2}.
Capturing the dewetting transitions of water under strong hydrophobic confinement and their coupling to the ionic dynamics would require a more detailed molecular modeling \cite{2005_Chandler, 2012_Hummer}.
However, our primary interest here is the description of the mesoscopic electrokinetic properties of the channel, which we show to be insensitive to the hydrodynamic characteristics of the hydrophobic constriction.
Although we will not be able to predict the electrolyte dynamics inside the stem area in atomic detail, this theoretical framework provides a
reliable description of the electrokinetic properties at the mesoscopic scale of the protein channel.
Nevertheless, solving the coupled Stokes-Poisson-Nernst-Planck equations in complex geometries has proven to be a challenging endeavor \cite{2010_Pagonabarraga}.

Here, we consider a model of a mechanosensitive channel consisting of the essential structural features found in \mscs{} and \mscl{}: a funnel-shaped pore with an uncharged hydrophobic stem and a vestibule carrying a fixed surface charge density, embedded in an impermeable membrane separating two solutions with salt concentrations $c_1$ and $c_2$, respectively (Fig. \ref{fig:boundary-conditions}C--D).
This model is based directly on the experimentally determined protein crystal structure, and aims to explain experimental work showing, first, that added charge in the vestibule activates the channel \cite{2001_Yoshimura,2004_Bartlett, 2006_Bartlett, 2002_Batiza, 2012_Doerner}, and second, that the hydrophobicity of the constriction provides a barrier for channel activation \cite{1999_Yoshimura}.
As an experimental benchmark, we consider measurements showing that \mscl{} and \mscs{} are activated at a hypotonic shock of at least $c_2 - c_1 = -0.3$~M \cite{1999_Levina}.

\textit{Governing equations.} --
We define a non-dimensional electrostatic potential $\psi\left(\myvec{x}\right) = e \phi\left(\myvec{x}\right)/(\kT{})$, with $\phi\left(\myvec{x}\right)$ being the potential in Volt and $\myvec{x} = \left(r,z\right)$ being the position in cylindrical coordinates.
The Poisson equation relates the electrostatic potential to the ion densities $c_{\pm}\left(\myvec{x}\right)$,
\begin{equation}
\nabla^2 \psi\left(\myvec{x}\right) = -4\pi b \left[c_{+}\left(\myvec{x}\right) - c_{-}\left(\myvec{x}\right)\right],
\label{eqn:poisson}
\end{equation}
with $b = e^2/\left(4\pi\varepsilon\varepsilon_0 \kT{}\right)$ being the Bjerrum length.
At low Reynolds number, the solvent velocity $\myvec{u}\left(\myvec{x}\right)$ is governed by the Stokes equation, which for an incompressible fluid in steady state reads
\begin{equation}
\nabla \cdot \left[\myten{P}\left(\myvec{x}\right) + \myten{T}\left(\myvec{x}\right)\right] + \myvec{f}\left(\myvec{x}\right) = 0 \quad \textrm{and} \quad \nabla\cdot\myvec{u}\left(\myvec{x}\right) = 0.
\label{eqn:stokes}
\end{equation}
The components of the viscous and electrostatic stress tensors $\myten{P}\left(\myvec{x}\right)$ and $\myten{T}\left(\myvec{x}\right)$ and the force density $\myvec{f}\left(\myvec{x}\right)$ due to the ionic \pmf{} $\mu\left(\myvec{x}\right)$ are given by
\begin{equation}
\begin{split}
\myten{P}_{ij}\left(\myvec{x}\right) &= -p\left(\myvec{x}\right)\delta_{ij} + \eta\left[\nabla_i u_j\left(\myvec{x}\right) + \nabla_j u_i\left(\myvec{x}\right)\right] \\
\myten{T}_{ij}\left(\myvec{x}\right) &= \frac{\kT{}}{8\pi b}\left[2 \nabla_i \psi\left(\myvec{x}\right) \nabla_j \psi\left(\myvec{x}\right) - \delta_{ij}\left(\nabla\psi\left(\myvec{x}\right)\right)^2\right] \\
\myvec{f}\left(\myvec{x}\right) &= -\kT{}\left[c_+\left(\myvec{x}\right)+c_-\left(\myvec{x}\right)\right] \nabla\mu\left(\myvec{x}\right),
\end{split}
\label{eqn:stress-tensor}
\end{equation}
with $p\left(\myvec{x}\right)$ being the hydrostatic pressure, $\eta$ being the viscosity and $i,j$ being $r,z$.
Inserting Eq. \ref{eqn:stress-tensor} into Eq. \ref{eqn:stokes} and taking the curl results in the following equations for the vorticity $\omega\left(\myvec{x}\right) = \nabla \times \myvec{u}\left(\myvec{x}\right) = \nabla_z u_r\left(\myvec{x}\right) - \nabla_r u_z\left(\myvec{x}\right)$,
\begin{equation}
\begin{split}
0 &= \eta \nabla^2\omega\left(\myvec{x}\right) + \nabla\times\left[\nabla\cdot\myten{T}\left(\myvec{x}\right) + \myvec{f}\left(\myvec{x}\right)\right] \\
\omega\left(\myvec{x}\right) &= r^{-1}\left(\nabla_z^2 + r \, \nabla_r \, r^{-1} \nabla_r\right)\xi\left(\myvec{x}\right).
\end{split}
\label{eqn:potential_flow}
\end{equation}
From the latter definition of $\xi\left(\myvec{x}\right)$ it follows
$u_r\left(\myvec{x}\right) = r^{-1}\nabla_{z} \, \xi\left(\myvec{x}\right)$ and $u_z\left(\myvec{x}\right) = -r^{-1}\nabla_{r} \, \xi\left(\myvec{x}\right)$,
which guarantees that the incompressibility condition is satisfied.
The local ion concentrations $c_{\pm}\left(\myvec{x}\right)$ are determined by conservation of species.
In steady state:
\begin{equation}
\myvec{u}\left(\myvec{x}\right) \cdot \nabla c_{\pm}\left(\myvec{x}\right) = -\nabla\cdot\myvec{J}_{\pm}\left(\myvec{x}\right),
\label{eqn:nernst-planck}
\end{equation}
with $\myvec{u}\left(\myvec{x}\right)$ being the velocity of the solvent, $c_{\pm}\left(\myvec{x}\right)$ the concentrations of positive and negative ions, and $\myvec{J}_{\pm}\left(\myvec{x}\right)$ the corresponding fluxes, given by
\begin{equation}
\myvec{J}_{\pm}\left(\myvec{x}\right) = -D_{\pm} \left[\nabla c_{\pm}\left(\myvec{x}\right) + c_{\pm}\left(\myvec{x}\right) \left(\nabla \mu\left(\myvec{x}\right) \pm \nabla \psi\left(\myvec{x}\right) \right) \right],
\label{eqn:flux}
\end{equation}
with $D_{\pm} = 1$~nm$^{2}$/ns being the ionic diffusion constant.
We numerically solve Eqs. \ref{eqn:poisson}--\ref{eqn:flux} in the domain shown in Fig. \ref{fig:boundary-conditions}D using a finite-difference over-relaxation scheme, which allows us to analyze the diffusio-osmotic force exerted on the channel wall for the first time.

\textit{Boundary conditions.} --
We employ a fixed surface charge density in the conical vestibule of $\sigma = -0.5$~$e\,$nm$^{-2}$ and uncharged boundaries everywhere else.
The hydrodynamic equations obey the no-slip boundary condition on the surface of the membrane and the vestibule, as is appropriate for hydrophilic surfaces \cite{2010_Bocquet_ChemSocRev, 2013_Bonthuis_JPCB}.
Inside the hydrophobic constriction, on the other hand, perfect slip is assumed, consistent with the plug-like flow found in hydrophobic nanotubes \cite{2009_Thomas-McGaughey}.
Note that assuming no slip inside the hydrophobic constriction instead leads to very similar results, implying that the model is robust regarding the characteristics of the hydrodynamic flow inside the constriction.
The normal flux vanishes at the membrane and pore boundaries, $\myvec{J}_{\pm} \cdot \hat{n} = 0$.
A fixed pressure difference between the open boundaries is achieved by adjusting the fluid flow through the box. 
Guided by experimental design, we set $c_1 = 0.5$~M \cite{1999_Levina}.
Molecular dynamics simulations show that the one dimensional ionic \pmf{} $\mu\left(z\right)$ in narrow channels exhibits a peak, reaching a maximum of $\mu_{0} \sim 10$--$30$~$\kT{}$ in the center of the channel, which decreases with increasing channel radius \cite{2004_Beckstein_JACS, 2004_Allen_PNAS, 1969_Parsegian, 2012_Richards, 2004_Anishkin}.
At a radius of $R = 1$~nm, $\mu_{0}$ is still several $\kT{}$'s in short nanopores \cite{2004_Beckstein_JACS}.
Therefore, we model the ionic \pmf{} by a repulsive potential in the stem of the funnel of height $\mu_{0}$, that decreases linearly from $\mu_0 = 18$~$\kT{}$ at $R = 0.3$~nm to zero at $R = 1.2$~nm.
This potential comprises all interactions between the ions, the water and the pore, including changes in the hydration state of the pore \cite{2012_Hummer}.

The force on the surface $\mathcal{S}$ of the pore, consisting of the stem and the vestibule, is calculated from the normal stress,
$\myvec{F}\left(R\right) = - \int_{\mathcal{S}} \left(\myten{P}\left(\myvec{x}\right) + \myten{T}\left(\myvec{x}\right)\right) \cdot \hat{n} \; \textrm{d}\myvec{x}$.
We calculate the nonequilibrium free energy landscape 
as the sum of two terms: the integral over the radial force $F_r\left(R\right)$ due to the electrolyte, and an elastic term due to the protein and the membrane,
\begin{equation}
G\left(R\right) = -\int_{R_0}^{R} F_r\left(R'\right) \textrm{d}R' + \pi K \left(R^2-R_0^2\right),
\label{eqn:free-energy}
\end{equation}
with $R_0 = 0.3$~nm being the minimum channel radius.
For the elasticity coefficient of the protein and the membrane we assume $K = 0.5$~$\kT{}\,$nm$^{-2}$, which is well within the range of values quoted in literature \cite{2009_Phillips_insight}.

\begin{figure}
\includegraphics[width=8cm]{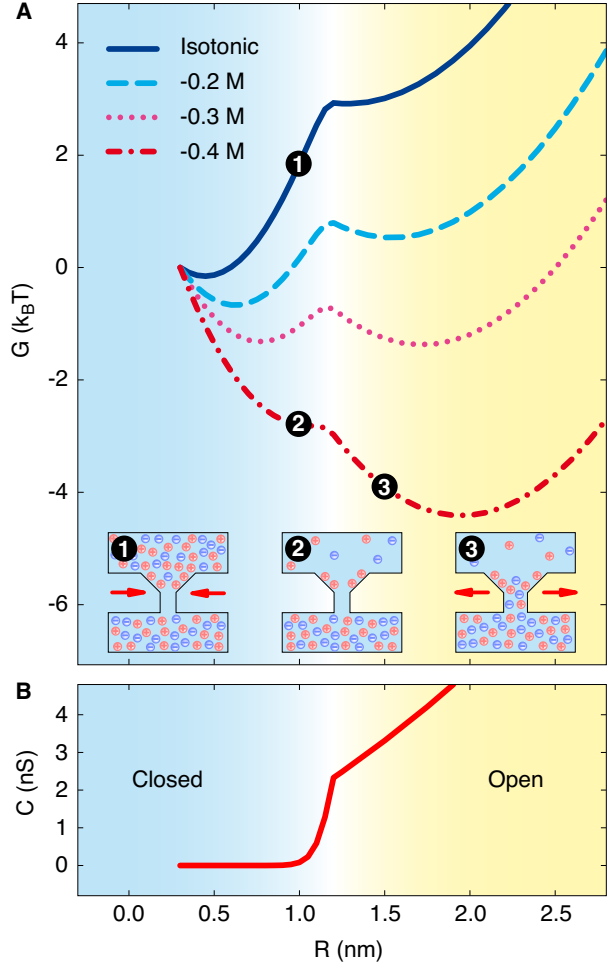}
\caption{
(A) Nonequilibrium free energy (Eq. \ref{eqn:free-energy}) as a function of the channel radius $R$ for different salt concentrations: isotonic ($c_1 = c_2 = 0.5$~M) and hypotonic ($c_2-c_1 = -0.2$~M, $-0.3$~M and $-0.4$~M).
For small radii, ions are excluded from the stem area by the ionic \pmf{}, leading to a contractile net force under isotonic conditions at $R = 1.0$~nm (inset 1).
After hypotonic shock, the increased electrostatic pressure diminishes the net force (inset 2).
At $R = 1.5$~nm, the ionic \pmf{} has vanished, ions enter the hydrophobic constriction and the channel activates under influence of the expansile electrostatic pressure (inset 3).
(B) Electrical conductance of the channel at a salt concentration of $c_1 = c_2 = 0.3$~M.
}
\label{fig:free-energy}
\end{figure}

Within this theoretical framework, the tension on the channel wall results from a competition of contractile forces due to the ionic \pmf{} and the elastic membrane, and expansile forces due to the charged vestibule.
The striking result of this competition is that the nonequilibrium free energy landscape $G\left(R\right)$ exhibits two minima, corresponding to the closed and open states (Fig. \ref{fig:free-energy}A).
Under isotonic conditions, exclusion of ions from the hydrophobic stem at small radii (inset 1 of Fig. \ref{fig:free-energy}A), which is known to reduce the pressure between like-charged parallel plates \cite{2004_Edwards-Williams}, gives rise to an energy barrier between the two states of $\sim 3$~$\kT{}$.
Remarkably, the energy barrier arises naturally from only electrostatic and hydrodynamic forces.
The second energy minimum is caused by the expansile electrostatic force, which increases upon hypotonic shock.
Whereas for $R < 1.2$~nm the increased electrostatic force is partly compensated for by the reduced pressure due to the ionic \pmf{} (inset 2 of Fig. \ref{fig:free-energy}A), 
the electrostatic force dominates when $\mu_{0} \rightarrow 0$ for $R > 1.2$~nm, and the channel activates (inset 3 of Fig. \ref{fig:free-energy}A).
For large $R$ the elastic term overcomes the electrostatic repulsion.
The first order transition between closed and open states occurs at a hypotonic shock of approximately $c_2-c_1 = -0.3$~M, 
in quantitative agreement with experimental results \cite{1999_Levina}.
The profiles show that the tension on the pore wall due to the electrolyte is sufficient to activate a mechanosensitive channel.

The channel activation is evident from the electrical conductance (Fig. \ref{fig:free-energy}B), which we calculate from $C\left(R\right) = \frac{\textrm{d}I\left(R\right)}{\textrm{d}\Delta\psi}$, with $\Delta\psi = \psi_2-\psi_1$ being an applied potential difference across the channel and 
$I\left(R\right) = e \int \myvec{J}_{+}\left(\myvec{x}\right) - \myvec{J}_{-}\left(\myvec{x}\right) + \myvec{u}\left(\myvec{x}\right)\left[c_{+}\left(\myvec{x}\right) - c_{-}\left(\myvec{x}\right)\right]\,\textrm{d}\myvec{x}$ 
being the resulting electric current, where the integration can be carried out over any plane spanning the pore.
The asymmetry in the conductance with respect to the direction of the applied potential difference, which is due to the asymmetric geometry of the channel, is negligible.
The conductance is minute up to a radius of $R = 1$~nm (Fig. \ref{fig:free-energy}B), owing to the repulsive \pmf{}.
Between $1.0 < R < 1.2$~nm, the conductance increases dramatically, before adopting linear growth with $R$.
To compare with experimental data, the salt concentration is set to $c_1 = c_2 = 0.3$~M.
The conductance of the open channel agrees well with the experimental values of $2.5-3.7$~nS measured for \mscl{} \cite{1997_Sukharev, 1994_Sukharev_Nature}.

\begin{figure}
\includegraphics[width=8cm]{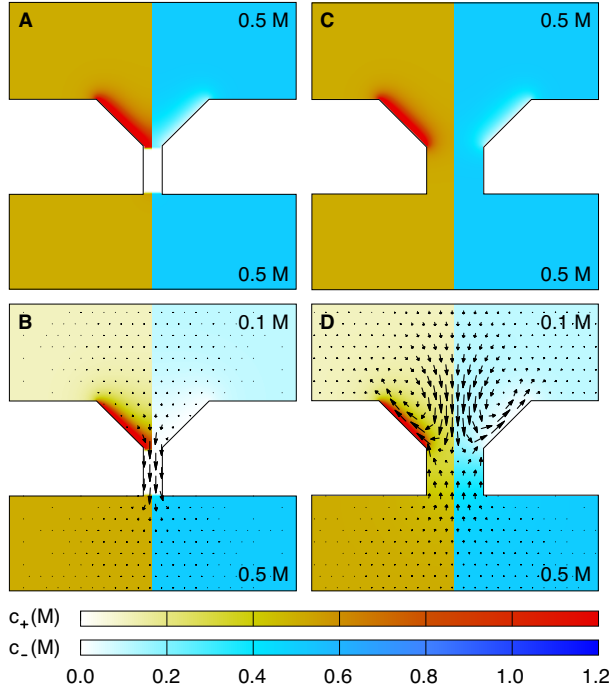}
\caption{
Ion concentrations $c_+\left(\myvec{x}\right)$ (left panels) and $c_-\left(\myvec{x}\right)$ (right panels) and fluid velocity (arrows) for channels in the closed ($R = 0.5$~nm) and open ($R = 1.5$~nm) states.
Under isotonic conditions, the fluid velocity is zero, and ions are repelled from the stem region in the closed state (A), but not in the open state (C).
Under hypotonic conditions, the fluid velocity is directed cell inward in the closed state (B) and outward in the open state (D).
The scale of the arrows in D has been increased by a factor 40 relative to those in B.
}
\label{fig:concentrations}
\end{figure}

To examine the functionality of the channel, we monitor the ion concentrations and water flux throughout the activation process.
In the closed state ($R=0.5$~nm), the ionic \pmf{} excludes both ion types from the stem of the funnel, as revealed by the concentrations $c_{\pm}\left(\myvec{x}\right)$
(Figs. \ref{fig:concentrations}A--B).
In the open state ($R=1.5$~nm), on the other hand, ions flow through the channel uninhibited (Figs. \ref{fig:concentrations}C--D).
In response to a hypotonic shock, water rushes into the cell, driven by the osmotic pressure (arrows in Fig. \ref{fig:concentrations}B).
When the channel activates, ions flowing outward through the channel drag the fluid along, and the water flow reverses (arrows in Fig. \ref{fig:concentrations}D), thus reproducing the experimentally observed behavior.

In conclusion, two-state mechanosensitive channel gating emerges from the electrokinetic transport equations without phenomenological assumptions in a simplified geometry that is based directly on the experimentally determined protein crystal structure.
Our proposed gating mechanism is fully supported by mutation experiments, which show a strong influence of protein surface charge and hydrophobicity on the gating kinetics.
Moreover, it agrees quantitatively with experiments regarding hypotonic shock threshold and electrical conductivity.
The activation mechanism can be verified further using mutation experiments, substituting charged residues for neutral ones.
Although there is evidence indicating that membrane-protein interactions also play a role in the gating transition, the direct response to hypotonic shock proposed in this work provides a faster and more accurate mechanism, bypassing the inhomogeneous cell membrane.
This novel modeling scheme reveals the underlying physics of the channel's complex biological function, showing that the gating kinetics can be fully reproduced within a model consisting of only a charged vestibule and a hydrophobic constriction.
Because these elements are shared features of many different ion channels, our proposed two-state gating mechanism is likely to be important for a broad range of ion channels.
Moreover, this new insight into the gating mechanism constitutes an essential step toward the design of artificial mechanosensitive channels and ion channel-targeting therapeutics.

\begin{acknowledgments}
We thank Philip Biggin and Julia Yeomans for valuable discussions of the manuscript.
D.J.B. acknowledges the Glasstone Benefaction and Linacre College for funding.
R.G. was supported by Human Frontier Science Program (HFSP) grant RGP0061/2013.
\end{acknowledgments}

\end{document}